\documentclass[12pt,aps,prd,preprint,tightenlines,superscriptaddress,
   showpacs,nofootinbib]{revtex4-1}
\usepackage{graphicx}

\usepackage{amsmath}
\usepackage{array}

\newcommand{\PRE}[1]{{#1}} 

\newcommand{\nbox}{{\,\lower0.9pt\vbox{\hrule \hbox{\vrule height 0.2 cm
\hskip 0.2 cm \vrule height 0.2 cm}\hrule}\,}}

\newcommand{\gev}{\text{GeV}}
\newcommand{\tev}{\text{TeV}}
\newcommand{\pb}{\text{pb}}

\newcommand{\cm}{\text{cm}}

\newcommand{\Dsl}[1]{\slash\hskip -0.20 cm #1}

\newcommand{\be}{\begin{equation}}
\newcommand{\ee}{\end{equation}}
\newcommand{\bea}{\begin{eqnarray}}
\newcommand{\eea}{\end{eqnarray}}
\newcommand{\baln}{\begin{align}}
\newcommand{\ealn}{\end{align}}
\newcommand{\lsim}{\lower.7ex\hbox{$\;\stackrel{\textstyle<}{\sim}\;$}}
\newcommand{\gsim}{\lower.7ex\hbox{$\;\stackrel{\textstyle>}{\sim}\;$}}

\usepackage{graphicx}
\begin{document}

\preprint{UH-511-1215-2013, CETUP2013-011}

\title{
\PRE{\vspace*{1.3in}}
Dipole Moment Bounds on Dark Matter Annihilation
\PRE{\vspace*{0.3in}}
}

\author{Keita Fukushima}
\affiliation{Department of Physics and Astronomy, University of
Hawai'i, Honolulu, HI 96822, USA
\PRE{\vspace*{.1in}}
}

\author{Jason Kumar
}
\affiliation{Department of Physics and Astronomy, University of
Hawai'i, Honolulu, HI 96822, USA
\PRE{\vspace*{.1in}}
}


\begin{abstract}
\PRE{\vspace*{.3in}}
We consider constraints on simplified models in which scalar dark matter annihilates to
light charged leptons through the exchange of charged mediators.  We find that
loop diagrams will contribute corrections to the magnetic and electric dipole
moments of the light charged leptons, and experimental constraints on these corrections
place significant bounds on the dark matter annihilation cross section.  In particular,
annihilation to electrons with an observable cross section would be ruled out, while
annihilation to muons is only permitted if the dominant contributions arise from
$CP$-violating interactions.
\end{abstract}


\maketitle

\section{Introduction} \label{sec:intro}

With the appearance of data from a variety of direct, indirect and collider
search strategies, there has been great interest in formalisms which allow one
to, for any particular model, relate the expected observables which can be probed
with each detection strategy.  One such formalism is that of simplified models~\cite{SimplifiedModels}, in
which one focusses on a effective Lagrangian involving only the small number
of fields and couplings relevant for dark matter interactions with the Standard Model.  In its
most basic form, a simplified model could describe dark matter interactions with
the Standard Model particles through
exchange of a single mediating particle.  For such a simplified model, one can describe
the full set of dark matter-Standard Model interactions by specifying only a
few quantities, such as the mass and spin of the dark matter and mediating particle,
the strength of the couplings, and the exchange channel ($s$, $t$ or $u$).  A single
simplified model can thus capture the essential physics of dark matter-Standard Model
interactions for a variety of new physics scenarios.

\begin{figure}[b]
\raisebox{-0.5\height}{ \includegraphics[scale=0.8]{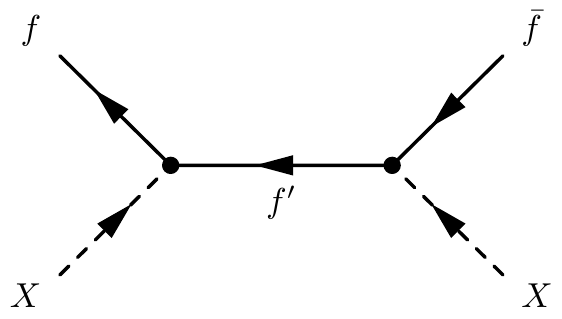} }\
\hspace{0.05 \textwidth}
\raisebox{-0.5\height}{ \includegraphics[scale=0.8]{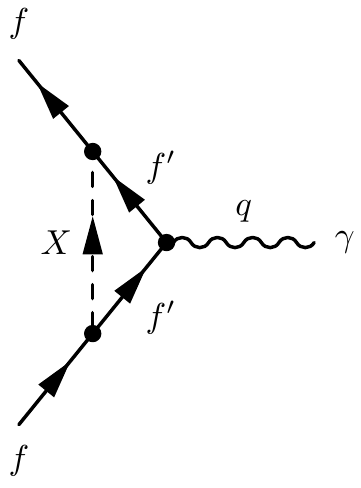} }
\label{fig:FeynmanDiagrams}
\caption{Feynman diagrams for the annihilation
process $XX \rightarrow \bar f f$ (left) and for the lepton dipole moment corrections (right).}
\end{figure}

We consider a class of simplified models in which dark matter is a scalar, and
dark matter annihilation to light charged leptons ($XX \rightarrow e^+ e^-, \mu^+ \mu^-$)
proceeds through $t$-channel exchange of a single
mediating particle, $f'$ (see fig.~\ref{fig:FeynmanDiagrams}, left panel).
Examples of models which can realize this scenario include WIMPless dark matter~\cite{WIMPless},
and this scenario can be realized in models of leptophilic dark matter~\cite{LeptophilicDarkMatter}.
This particular scenario is of interest because it allows one to decouple the dark matter annihilation
cross section from bounds arising from direct detection and collider searches.
These latter search strategies place tight constraints on dark matter interactions with
quarks.  In some (though by no means all) models, these bounds are satisfied by requiring the
particles mediating interactions with quarks to be very
heavy, suppressing the annihilation cross section to quarks.  But if the particles mediating
interactions with leptons are light, then the cross section for dark matter to annihilate
to leptons could still be sizable.  In this light, it is important to consider the case
where dark matter is a scalar whose annihilation cross section does not suffer from the
chirality/$p$-wave suppression which typically arises for Majorana fermions.

Because the dark matter has vanishing electric charge,
the mediating particle must necessarily be electrically charged.  As a result,
the dark matter-lepton interaction vertices can contribute to the electric or
magnetic dipole moments of the electron and muon (see fig.~\ref{fig:FeynmanDiagrams}, right panel).
Both electric and magnetic dipole moments have long been used a probes of new physics (see, for
example,~\cite{NewPhysics,ElectronMagMoment,MuonMagMoment,Roberts:2011fq,ElecMoment}).
In particular, the current measurement of the muon magnetic dipole moment differs from the Standard
Model prediction by $\sim 3.5\sigma$~\cite{MuonMomentAnomaly}, and there have been attempts to
interpret this anomaly in terms of a signal for new physics~\cite{MuonMagMomentNewPhysics}.
We will consider the connection between corrections to the dipole moments and the dark matter
annihilation cross section.

We will consider the case where dark matter is a scalar, with the most general
allowed Yukawa coupling to light charged leptons and an exotic
charged lepton.  We will find that the annihilation cross section depends on
two terms, one of which is $CP$-conserving and the other of which is $CP$-violating.
The $CP$-conserving term will be constrained by the new physics contribution to the
magnetic dipole moment of the charged lepton, while the $CP$-violating term will constrained
by the electric dipole moment.  This will allow us to constrain the the $XX \rightarrow \bar f f$
annihilation cross-section with experimental measurements of the fermion dipole moments.
We thus focus on annihilation to light charged leptons (electrons or muons), whose dipole moments
are most tightly constrained.

The outline of this paper is as follows.  In section II, we describe the simplified
model, and the scaling of $CP$-conserving and $CP$-violating terms.  In section III we
compute the $XX \rightarrow \bar f f$ annihilation cross section and the corrections
to the electric and magnetic dipole moments.  In section IV, we describe experimental
constraints on the dipole moments, and the implied constraints on the dark matter annihilation
cross section.  We conclude with a discussion of our results in section V.

\section{The Simplified Model}

If dark matter is a scalar, which couples to Standard Model leptons and an
exotic fermion, then the most generic renormalizable interaction can be
written in terms of the effective Lagrangian:
\bea
L_{X} =  X \bar f' (\lambda_L P_L + \lambda_R P_R) f  +   X^* \bar f (\lambda_L^* P_R + \lambda_R^* P_L) f' ,
\eea
where $f$ is the Standard Model lepton, $f'$ is an exotic lepton, and $X$ is a scalar dark matter field, which
in general can be real or complex.  We also assume that $X$ and $f'$ are charged under
some unbroken symmetry which stabilizes the dark matter, and under which $f$ is neutral.  We thus require
$m_X < m_{f'}$.

If $X$ is an $SU(2)_L \times U(1)_Y$ singlet, then gauge-invariance implies that
$f'$ must be chiral under $SU(2)_L$ and $U(1)_Y$.  As a result, $f'$ will act like an exotic lepton and will get mass
through a coupling to the higgs; it cannot be arbitrarily
heavy, though it is certainly allowed by data.  Note, however, that if $f'$ is part of a full generation (to
cancel the hypercharge mixed anomaly), then this would imply the existence of a 4th generation quark, which is
more tightly constrained~\cite{4thGenQuarks}.  However, the anomaly can be cancelled in other ways (say, the presence of another
mirror lepton).  For our purposes, we simply assume that the hypercharge mixed anomaly is cancelled.

Note that it is also possible for $f'$ to be completely vector-like, in which case its mass is unconstrained.
But in this case, gauge-invariance implies that $X$ must be a linear combination of fields with different
charges under $SU(2)_L$ and $U(1)_Y$ (though electrically neutral).  The coefficients $\lambda_{L,R}$ then
also include the relevant mixing angles.
We will thus treat the mass of $f'$ as unconstrained (aside from $m_{f'}>m_X$), though for specific models
$m_{f'}$ can be constrained both from above and from below by data.

We can choose to write the propagators for $f$ and $f'$ with real mass terms.  With this choice, the only remaining
allowed phase rotation of the fields is an overall rotation of $X$, $f$ and $f'$ (the relative phase between
the left-handed and right-handed fermion components are fixed by the real mass condition).  Any
overall phase for the coefficients $\lambda_L$ and $\lambda_R$ can be absorbed by a phase rotation of the fields,
but a relative phase between $\lambda_L$ and $\lambda_R$ cannot be absorbed by any field redefinition.  We thus
see that $CP$-violating
terms will be proportional to $Im(\lambda_L \lambda_R^*)$.

\section{Connecting Dark Matter Annihilation to Electric/Magnetic Dipole Moments}

\subsection{Dark Matter Annihilation}

If $X$ is real, then the $XX \rightarrow \bar f f$ annihilation process proceeds
though $t$- and $u-$channel exchange of $f'$.  If $X$ is complex, then the $X^* X \rightarrow \bar f f$
annihilation process proceeds through only one of these channels.

For scalar dark matter, the $XX \rightarrow \bar f f$ process can proceed from an $s$-wave initial state,
and with no chirality suppression.  As a result, although velocity-dependent terms can be relevant for
the annihilation rate at freeze-out, they will be largely irrelevant in the current epoch.  We can thus
limit ourselves to the the non-relativistic limit.  We will also focus on the limit $m_X ,m_{f'} \gg
m_f$.\footnote{However, if either $\lambda_L$ or $\lambda_R$ is sufficiently small, then the leading
terms in the cross section will scale as $m_f^2$ or $v^4$~\cite{3body}.  Since these terms will be small,
internal bremsstrahlung can also
be important.  The annihilation cross section is suppressed in this limit, so we will not
discuss it further.    }
We then find
\bea
\sigma(XX \rightarrow \bar f f) v &\simeq &  \left[ Re(\lambda_L \lambda_R^*)^2 + Im(\lambda_L \lambda_R^*)^2 \right]
 {  m_{f'}^2\over \pi  ( m_{f'}^2 + m_{X}^2)^2  } ,
\nonumber\\
\sigma(X^* X \rightarrow \bar f f) v &\simeq &  \left[ Re(\lambda_L \lambda_R^*)^2 + Im(\lambda_L \lambda_R^*)^2 \right]
 {  m_{f'}^2\over 4\pi  ( m_{f'}^2 + m_{X}^2)^2  } ,
\eea
as $v \rightarrow 0$.

It is useful to consider the limit $m_{f'} \gg m_X \gg m_f$.
In this limit, the annihilation matrix element
is generated by the dimension 5 effective operator:
\bea
{\cal O}_{eff} &=& {Re (\lambda_L \lambda_R^*) \over m_{f'}} (X^* X )(\bar f f)
- {Im (\lambda_L \lambda_R^*) \over m_{f'}} (X^* X)(\imath \bar f  \gamma^5 f) .
\eea
Note that each of the two terms permits annihilation from an $S=0$, $L=0$, $J=0$ $CP$-even initial state.  However, the
first term annihilates to a $S=1$, $L=1$ $CP$-even final state, while the second term annihilates to a
$S=0$, $L=0$ $CP$-odd final state.  The annihilation matrix elements generated by these two terms thus do not interfere,
and each matrix element is neither chirality nor $p$-wave suppressed~\cite{Kumar:2013iva}.

As expected, the terms in the cross section proportional to $Re (\lambda_L \lambda_R^*)$ arise from
the $CP$-invariant part of the operator, while the terms proportional to $Im (\lambda_L \lambda_R^*)$ arise
from the $CP$-violating part.

\subsection{Corrections to the Electric/Magnetic Dipole Moment}

The most general fermion-fermion-photon coupling consistent with gauge-invariance is
given by
\bea
\Gamma^\mu &= \gamma^\mu F_1 (q^2) +{\imath \sigma^{\mu \nu} q_\nu \over 2m} F_2 (q^2)
+{\imath \sigma^{\mu \nu} q_\nu \gamma^5  \over 2m} F_3 (q^2)
+(\gamma^\mu q^2 - \Dsl q q^\mu) \gamma^5 F_A (q^2) ,
\eea
where $F_{1,2,3,A}(q^2)$ are form factors which depend on the momentum $q$ of the photon.
Gauge-invariance requires $F_1 (0) =1$.  $F_A (0)$ is the anapole moment, which will
not concern us here.  But the electric and magnetic dipole moments are determined by
$-\imath F_3 (0)$ and $F_2 (0)$.  These moments are denoted by $d$ and $\mu = g (e /2m)S$,
respectively.  At the classical level, one finds a vanishing electric dipole moment, $d_{cl}=0$,
and a magnetic dipole moment given by $g_{cl}=2$.  The one-loop quantum corrections to these
dipoles are given by the quantities
\bea
a \equiv {g-2 \over 2} &=& F_2 (0),
\nonumber\\
{d \over |e|} &=& {\imath F_3(0) \over 2m}.
\eea

These corrections can be computed at lowest order from the one-loop correction to the
vertex.  Note that, at lowest order, neither correction is affected by the fermion wavefunction
renormalization, so a computation of the vertex correction is sufficient.
In the limit $m_f \ll m_X, m_{f'}$, we then find (see also~\cite{Cheung:2009fc})
\bea
\Delta a &=& {Re( \lambda_L \lambda_R^*  ) \over (4 \pi)^2 }
{m_{f}m_{f'}\left( (m_{f'}^2-m_X^2)(m_{f'}^2-3m_X^2) + 2 m_X^4 \log{m_{f'}^2 \over m_X^2}  \right) \over (m_{f'}^2-m_X^2)^3} ,
\nonumber\\
{d \over |e|} &=&  {Im(\lambda_L \lambda_R^*  ) \over 2(4 \pi)^2 }
{m_{f'}\left( (m_{f'}^2-m_X^2)(m_{f'}^2-3m_X^2) + 2 m_X^4 \log{m_{f'}^2 \over m_X^2}  \right) \over (m_{f'}^2-m_X^2)^3} .
\eea
As expected, the correction to the magnetic dipole moment depends on $Re(\lambda_L \lambda_R^*)$, while the
electric dipole moment, which arises from $CP$-violation, depends on $Im(\lambda_L \lambda_R^*)$.  Note that the dipole
moment corrections do not depend on whether $X$ is a real or complex scalar.

It is also worth noting that, in general, the dipole moment operators will run under renormalization group (RG) flow
between the energy scale at which one integrates out the $X$ and $f'$ fields ($\sim m_X$) and the energy scale
relevant for the vertex corrections ($\sim m_f$).  However,
since the $\lambda_{L,R}$ we consider are typically small and since the only other relevant couplings are electroweak couplings,
the effects of running are not significant~\cite{Jung:2008it} and we will ignore them.  If we had considered the coupling of dark matter to
quarks instead, RG-running of the operators would be non-trivial.

\section{Constraints and Prospects}

We focus on dark matter couplings to electrons or muons, whose dipole moments
are most tightly constrained.
We can compare the experimental measurement and theoretical calculation of $a=(g-2)/2$~\cite{muon,mm,electron}
for the electron and muon:
\bea
a_{e(exp)} = && 1159652180.76 (0.27) \times 10^{-12} ,
\nonumber\\
a_{e(theory)} = && 1159652181.13 (0.11)(0.37)(0.77) \times 10^{-12} ,
\nonumber\\
a_{e (exp)} - a_{e(theory)} = \Delta a_e = && -0.37 (0.82) \times 10^{-12} ;
\\
a_{\mu (exp)} = && 116592089.(54)(33) \times 10^{-11} ,
\nonumber\\
a_{\mu (SM)}  = && 116591802(2)(42)(26) \times 10^{-11} ,
\nonumber\\
 a_{\mu (exp)}  - a_{\mu (SM)} = \Delta a_{\mu}	= && 287(63)(49) \times 10^{-11} .
\eea

The Standard Model prediction for the electric dipole moment $d$ of the electron and
muon is negligible.
The experimental measurement~\cite{electron,Moyotl:2011yv} is:
\bea
\left| {d_{\text{e}} \over e} \right| &<& 10.5 \times 10^{-28}  \cm ,
\nonumber\\
2 m_e \left| {d_{\text{e}} \over e} \right| &<& 5.5 \times 10^{-17} ;
\\
\left| {d_\mu \over e} \right| &< & 1.5 \times 10^{-19}  \cm ,
\nonumber\\
2 m_\mu \left| {d_\mu \over e} \right| &< & 1.6 \times 10^{-6} .
\eea

We see that constraints on the corrections to $\Delta a$ and $d$ imply constraints
on $Re(\lambda_L \lambda_R^*)$ and $Im(\lambda_L \lambda_R^*)$, which in turn imply a bound on
the annihilation cross section.
But since there can be additional new physics which contribute to both $\Delta a$ and $d$, a contribution
to these moments from dark matter interactions alone which exceeds the difference between theory and the measured
value cannot be strictly excluded.
Instead, we will consider as ``disfavored" a model for which the magnitude of the correction from dark matter interactions
is larger than
the magnitude of the total correction allowed by the data (regardless of sign).  Although such models can be allowed
by data if there is additional new physics, those contributions would necessarily have to be fine-tuned against the
dark matter contribution.
Clearly, this characterization is somewhat ambiguous, and there is no clear answer as to how much fine-tuning is
acceptable; this analysis merely provides a benchmark.

\begin{figure}[h]
\includegraphics*[width=3in]{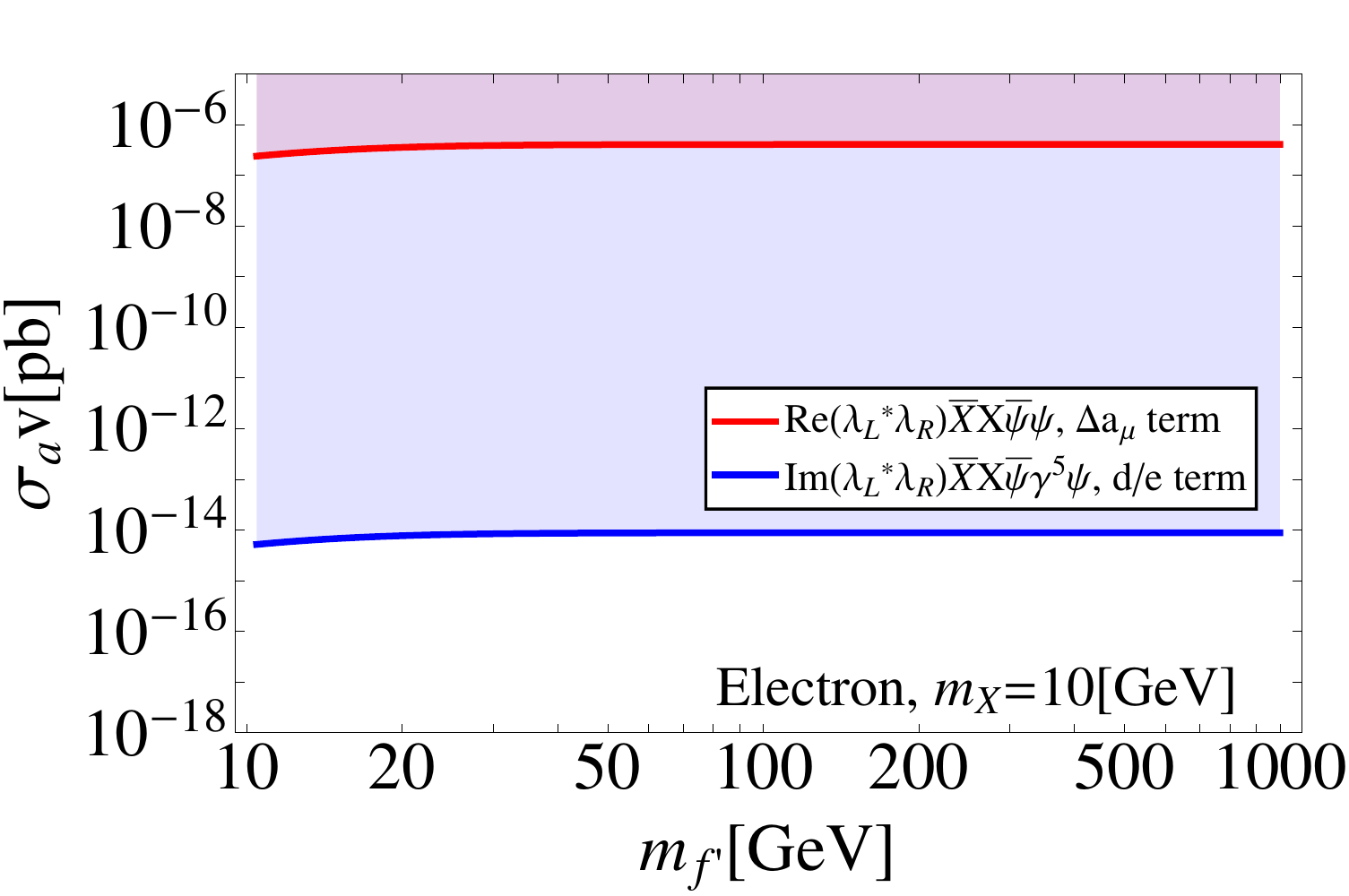}
\includegraphics[width=3in]{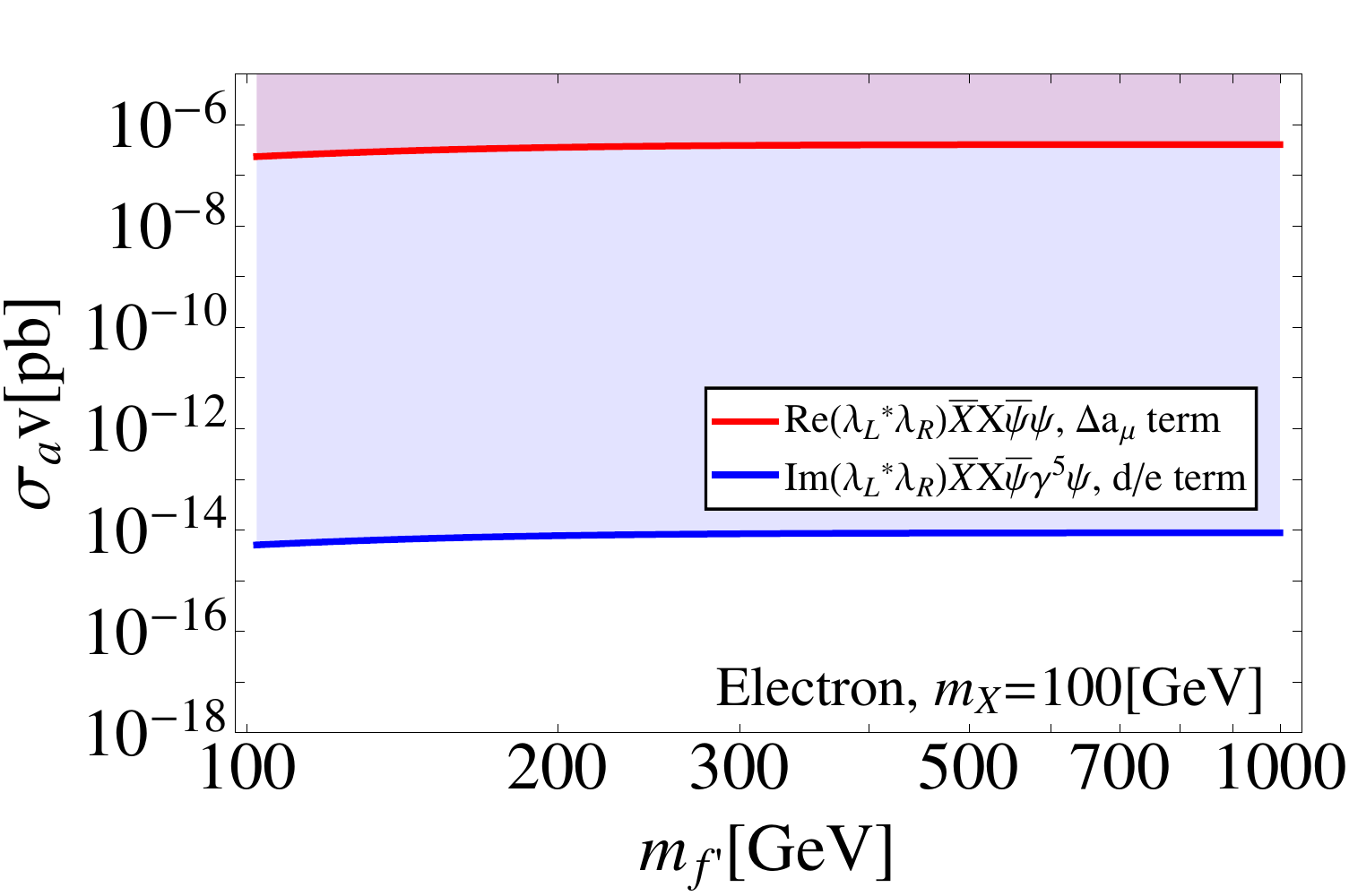}
\includegraphics*[width=3in]{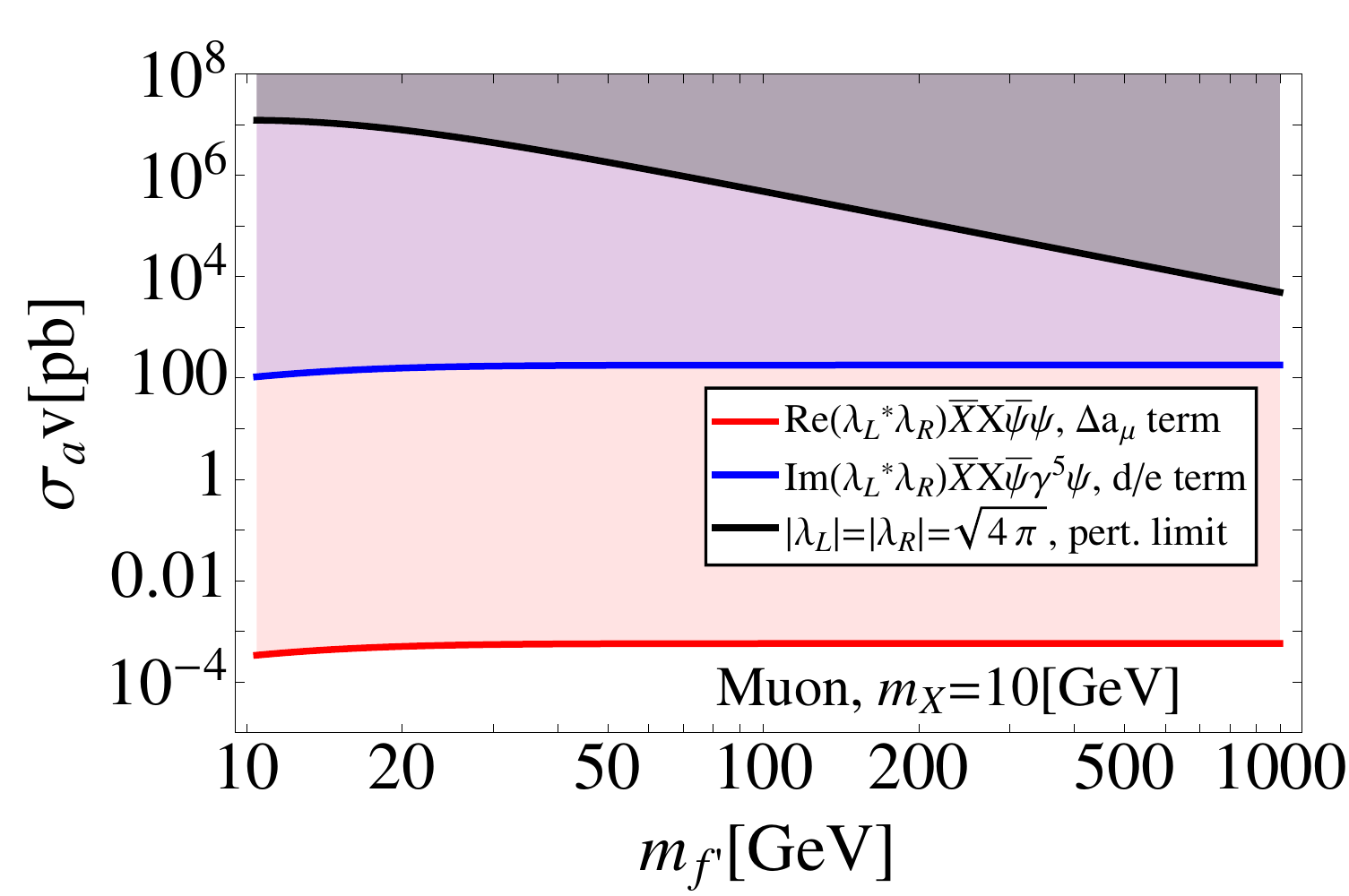}
\includegraphics[width=3in]{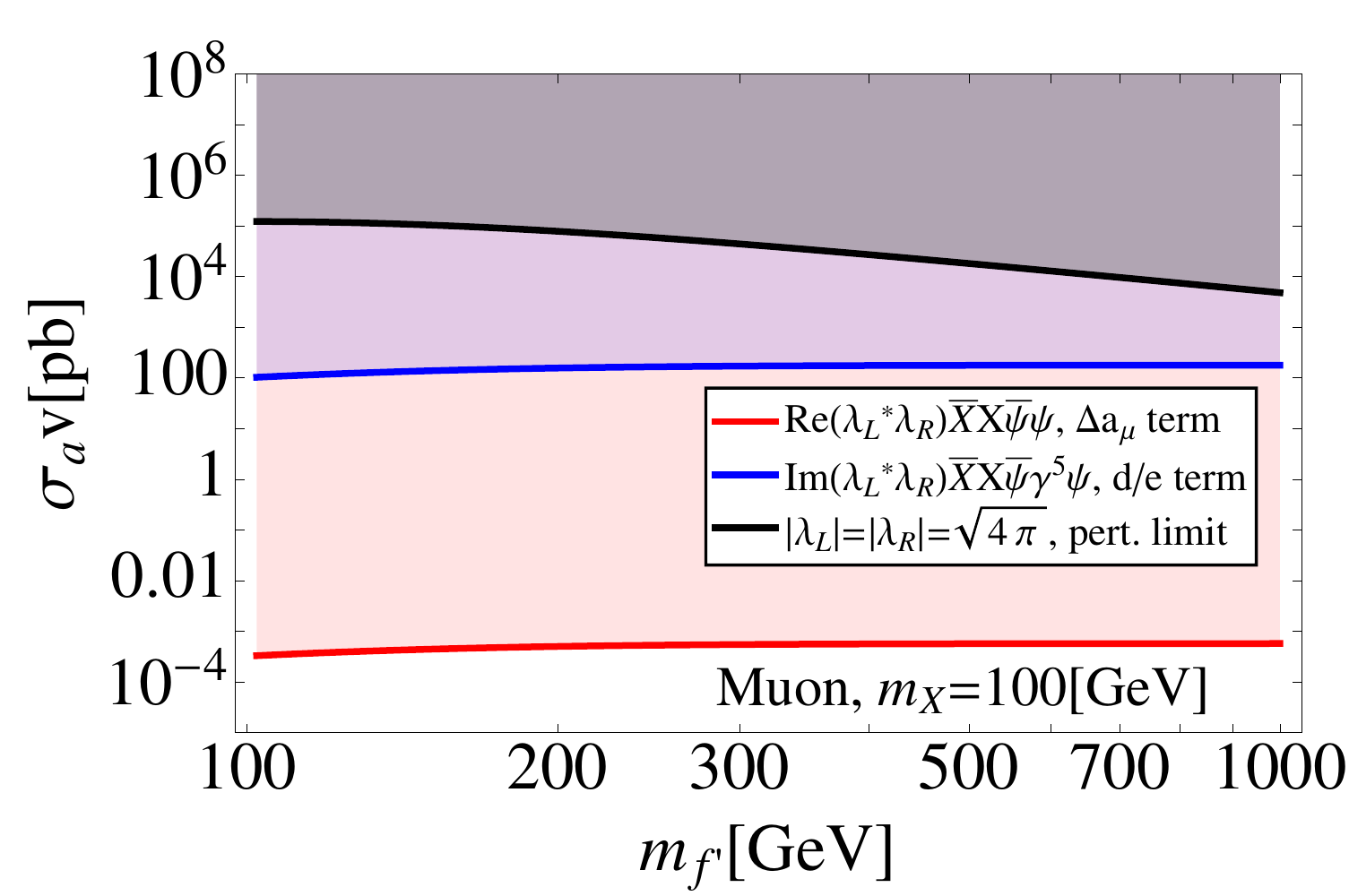}
\caption{Bounds on the annihilation cross section of complex scalar dark matter to $e^+ e^-$ (top panels)
and $\mu^+ \mu^-$ (bottom panels) from experimental
measurement of the lepton magnetic and electric dipole moments in the $m_f \to 0$
limit for $m_X =10~\gev$ (left panels) and  $m_X =100~\gev$ (right panels). Each disfavored region is
shaded and demarcated by a line
which indicates if it is disfavored by magnetic dipole moment bounds, electric dipole moment bounds, or by
perturbativity of the coupling constant (perturbativity is not constraining for $X^*X \rightarrow e^+ e^-$).
Note that the annihilation cross section will be enhanced by a factor of 4
if $X$ is a real scalar.  }
\label{fig:CrossSecBound}
\end{figure}

Using the analysis above, we plot in fig.~\ref{fig:CrossSecBound} regions of parameter-space in the
$(\langle \sigma_{a}v \rangle, m_{f'} )$ plane which are disfavored by the dipole moment data,
for $m_X = 10, 100~\gev$ and $m_f \ll m_X$.  The annihilation processes considered are $X^* X \rightarrow e^+ e^-$
and $X^* X \rightarrow \mu^+ \mu^-$.  In each panel the labeled shaded regions correspond to annihilation cross sections due either to
$CP$-conserving or $CP$-violating interactions, and are constrained by the appropriate
dipole moment.  Note that $m_{f'}$ is constrained by direct searches for charged particles at
LEP.  The precise exclusion contours depend on the spin of the new particle as well as its decay chain, but LEP searches
roughly exclude new charged particles which are
lighter than $\sim 100~\gev$~\cite{pdg,Carpenter:2011wb}.
LHC constraints on new charged leptons can be up to a factor of 3 tighter if the charged
lepton couples to $SU(2)_L$, but can be much weaker if they do not~\cite{lhcslepton,lhcsleptonrh}.  As these constraints are somewhat
model-dependent, we plot the full range of $m_{f'}$ for completeness.

It is interesting to note that, for annihilation to the electron channel, the bounds on both $CP$-conserving
and $CP$-violating contributions to the annihilation cross section are tight enough to exclude any models which
could be observed at current or upcoming indirect detection experiments, although bounds on $CP$-violating
contributions are tighter than bounds on the $CP$-conserving contribution.  On the other hand, for annihilation
to the muon channel, bounds from $g_\mu-2$ on the $CP$-conserving contribution are much tighter than
electric dipole moment bounds on the $CP$-violating contribution.  Indeed, if this simplified model describes
$X^* X \rightarrow \mu^+ \mu^-$ annihilation which can be detected at indirect detection experiments, it must
arise from $CP$-violating contributions.

One can compare the annihilation cross sections and dipole moment corrections more directly in the limit
$m_{f'} \gg m_X \gg m_f$.  In this limit, we find
\bea
\Delta a_f &\simeq & {Re( \lambda_L \lambda_R^*  ) \over (4 \pi)^2 } {m_f \over m_{f'}}
\nonumber\\
2m_f (d_f /|e|) &\simeq & {Im( \lambda_L \lambda_R^*  ) \over (4 \pi)^2 } {m_f \over m_{f'}}
\nonumber\\
\sigma(X^* X \rightarrow \bar f f ) v &\simeq &  \left[ Re(\lambda_L \lambda_R^*)^2 +  Im(\lambda_L \lambda_R^*)^2  \right]
{  1\over 4 \pi  m_{f'}^2   } \sim 64\pi^3 [\Delta a_f^2+ (2m_f (d_f /|e|))^2] m_{f}^{-2}
\nonumber\\
&\simeq & (7.7 \times 10^{11}~{\rm pb} )[\Delta a_f^2 + (2m_f (d_f /|e|))^2]\left({m_f \over {\rm GeV}} \right)^{-2}
\eea
In this limit, the annihilation cross section depends on $m_X$ and $m_{f'}$ only through the dipole moment
corrections.
As expected, we see that for $f=e$, the $CP$-violating contribution is more highly
suppressed than the $CP$-conserving contribution.  On the other hand, for $f=\mu$,
it is the $CP$-conserving contribution which is more highly suppressed.
With $\Delta a_\mu \sim 3 \times 10^{-9}$ and $d_\mu =0$ ($Im (\lambda_L \lambda_R^*)=0$),
we find $\sigma(X^*X \rightarrow \bar \mu \mu ) v \sim 6\times 10^{-4}~{\rm pb}$.  This implies that
a $CP$-conserving interaction producing a $g_\mu-2$ correction large enough to explain the
data would still produce a annihilation cross section of only $\sim 10^{-3}~\pb$.  If a larger
cross section is observed in indirect detection experiments, then there must be additional fine-tuning
between corrections to the muon magnetic dipole moment, or there must be $CP$-violating interactions
which also contribute to the muon electric dipole moment.

Note that, if this simplified model describes
the physics underlying the $g_\mu -2$ anomaly, $m_{f'}$ could be as large as $\sim 35~\tev$ while still
having perturbative couplings.  For the case of annihilation to electrons, perturbativity constraints
are not relevant for any of the mass scales considered.

\section{Conclusions}
\label{sec:conclusions}

We have studied dipole moment constraints on simplified models in which scalar dark matter
annihilates to light charged leptons through $t$-channel exchange of a heavy
mediating fermion.  In particular, we have found that such simplified models are tightly
constrained by the magnitude of experimentally allowed corrections to electric and magnetic
dipole moments.  Interestingly, the tightest constraints on electron interactions arise from
electric dipole moment measurements, while the tightest constraints on muon interactions arise
from magnetic dipole moment interactions.  If such a simplified model describes dark matter
interactions in nature, then an observable $XX \rightarrow \mu^+ \mu^-$ annihilation cross-section
must be accompanied by large $CP$-violation.   It is also worth noting that, if dark matter is a
thermal relic, then the $XX \rightarrow e^+ e^-$ process can be only a small part of total annihilation
at freeze-out, while $XX \rightarrow \mu^+ \mu^-$ can be the dominant annihilation process at freeze-out.
These bounds can be weakened if there are other new physics effects whose contribution to the
dipole moments cancels that from loop diagrams involving dark matter.  However, models with
cross sections far larger than these constraints can only be consistent if there is significant fine-tuning
between dipole moment corrections.

In the limit $m_{f'} \gg m_X \gg m_f$, both $\Delta a$  and
$m_f (d/e)$ scale as $m_f / m_{f'}$.
One could instead consider the case where $X$ is a fermion and the $t$-channel mediator $f'$
is a scalar.  This simplified model would again yield corrections to the electric and magnetic dipole
moments, but these moments would instead scale as
$(m_f m_X / m_{f'}^2) \times [(m_f / m_X)\,{\rm or}\,(\sin \alpha)]$, where $\alpha$ is the
scalar mixing angle~\cite{Cheung:2009fc}.
This suppression is due to the fact that dipole moment interactions require a helicity-flip for the
Standard Model lepton, which can be provided by the fermion mass term or by scalar mixing.
However, if $X$ is a fermion, then the $s$-wave annihilation cross section is suppressed
by a factor $m_X^2 / m_{f'}^2$  (if dark matter is a Majorana fermion and there is negligible scalar mixing,
the annihilation cross section is suppressed by an additional factor $m_f^2 / m_X^2$).  As a result, the bounds
on dark matter annihilation to light charged leptons for the case of fermionic dark matter  with
large scalar mixing may be expected to be roughly the same as those shown here for the scalar dark matter case.
The same is true if dark matter is a Majorana fermion, and there is negligible scalar mixing.
But if dark matter is a
Dirac fermion with negligible scalar mixing, then the constraints on dark matter annihilation from
dipole moments would be much weaker~\cite{Buckley:2013sca}.

We can thus compare the analysis of the simplified model considered here with that of an MSSM neutralino.  The correction
to the muon magnetic moment due to loop diagrams has been well studied~\cite{Cheung:2009fc},
and for the reasons stated above,
we expect the bounds on the cross section for $\chi \chi \rightarrow \mu^+ \mu^-$ to be similar to those
found here in the $CP$-conserving case, up to ${\cal O}(1)$ factors.  This amounts to the statement that,
even if superparticle corrections are large enough to be responsible for the deviation between the measured
$g_\mu -2$ and the Standard Model prediction, the branching fraction for dark matter annihilation to muons
in the current epoch will still be quite small.  This is not surprising, given that the neutralino is a Majorana
fermion.  One might have thought that a larger annihilation cross section would be allowed for scalar dark matter,
since the required helicity-flip is provided by the mediating fermion.  However, the mediating fermion provides
the same helicity-flip for the magnetic moment diagram, enhancing the correction to $g_\mu-2$.  Thus a $CP$-conserving
interaction involving scalar dark matter which provides the needed correction to $g_\mu-2$ would still yield
an annihilation cross section to muons which is very small.  But for a fixed coupling, the
mediating particle mass required to match the $g_\mu -2$ anomaly in the scalar dark matter case
will be much larger than in the neutralino dark matter case.
Moreover, large $CP$-violation is possible in the simplified model considered
here, allowing significant annihilation to muons which is unconstrained by $g_\mu-2$ bounds.

For the case of Dirac neutralinos, dark matter can annihilate from a $S=1$, $L=0$, $J=1$ initial state.  In this
case, no helicity-flip is needed for the final state.  A helicity-flip is still needed for the correction to $g_\mu -2$,
implying that the correction to $g_\mu -2$ can be quite small even if the cross section for annihilation to muons is
large~\cite{Buckley:2013sca}.

As we have seen, for Majorana fermion dark matter, the spin-flip required for an $s$-wave annihilation
cross section is correlated with the spin-flip needed for a contribution to the dipole moments.  But it
is worth noting that even dark matter with $p$-wave suppressed annihilation can still have the correct
thermal relic density, because a $p$-wave annihilation cross section is only suppressed by a factor $\sim 10$
at the time of freeze-out.
As a result, Majorana fermion dark matter with largely $p$-wave annihilation could arise as a thermal relic,
yet be unconstrained by dipole moment bounds.

In this work, we have only focussed on annihilation to light charged leptons.  Annihilation to
$\bar \tau \tau$ or $\bar q q$ can also be constrained, but experimental constraints on the dipole
moments are much weaker.  It would be interesting to revisit constraints on annihilation to those
channels from dipole moment corrections.  Finally, it is worth noting that all of these bounds on
the annihilation cross section can be avoided by models with $s$-channel annihilation, where the mediating
particle is neutral.  The observation
of dark matter annihilation to $e^+ e^-$ pairs would then provide interesting information regarding
the nature of the mediating particle.

\vskip .2in
\textbf{Acknowledgments}

We are grateful to M.~Buckley, D.~Hooper and X.~Tata for useful discussions.
J.~K.~thanks the Center for Theoretical Underground Physics and
Related Areas (CETUP* 2013) in South Dakota for their support and
hospitality during the completion of this work.
The work of J.~K.~before June 1, 2013 is supported in part by Department of Energy
grant DE-FG02-04ER41291.  The work of J.~K.~after June 15, 2013 is supported in
part by NSF CAREER Award PHY-1250573.


\begin{thebibliography}{99}

\bibitem{SimplifiedModels}
  D.~Alves {\it et al.}  [LHC New Physics Working Group Collaboration],
  J.\ Phys.\ G {\bf 39}, 105005 (2012)
  [arXiv:1105.2838 [hep-ph]].


\bibitem{WIMPless}
  J.~L.~Feng and J.~Kumar,
  Phys.\ Rev.\ Lett.\  {\bf 101}, 231301 (2008)
  [arXiv:0803.4196 [hep-ph]];
  J.~L.~Feng, J.~Kumar and L.~E.~Strigari,
  Phys.\ Lett.\ B {\bf 670}, 37 (2008)
  [arXiv:0806.3746 [hep-ph]];
  J.~L.~Feng, H.~Tu and H.~-B.~Yu,
  JCAP {\bf 0810}, 043 (2008)
  [arXiv:0808.2318 [hep-ph]].


\bibitem{LeptophilicDarkMatter}
  N.~Arkani-Hamed, D.~P.~Finkbeiner, T.~R.~Slatyer and N.~Weiner,
  Phys.\ Rev.\ D {\bf 79}, 015014 (2009)
  [arXiv:0810.0713 [hep-ph]];
  P.~J.~Fox and E.~Poppitz,
  Phys.\ Rev.\ D {\bf 79}, 083528 (2009)
  [arXiv:0811.0399 [hep-ph]].


\bibitem{NewPhysics}
  T.~Moroi,
  Phys.\ Rev.\ D {\bf 53}, 6565 (1996)
  [Erratum-ibid.\ D {\bf 56}, 4424 (1997)]
  [hep-ph/9512396];
  G.~C.~McLaughlin and J.~N.~Ng,
  Phys.\ Lett.\ B {\bf 493}, 88 (2000)
  [hep-ph/0008209];
  S.~R.~Choudhury, A.~S.~Cornell, A.~Deandrea, N.~Gaur and A.~Goyal,
  Phys.\ Rev.\ D {\bf 75}, 055011 (2007)
  [hep-ph/0612327].


\bibitem{MuonMagMoment}
  A.~Czarnecki and W.~J.~Marciano,
  (Advanced series on directions in high energy physics. 20);
  D.~Stockinger,
  (Advanced series on directions in high energy physics. 20).



\bibitem{ElectronMagMoment}
  G.~F.~Giudice, P.~Paradisi and M.~Passera,
  JHEP {\bf 1211}, 113 (2012)
  [arXiv:1208.6583 [hep-ph]].

\bibitem{Roberts:2011fq}
  B.~L.~Roberts,
  J.\ Phys.\ Conf.\ Ser.\  {\bf 295}, 012027 (2011)
  [arXiv:1101.2251 [hep-ex]].


\bibitem{ElecMoment}
  M.~Pospelov and A.~Ritz,
  (Advanced series on directions in high energy physics. 20).

\bibitem{MuonMomentAnomaly}
  G.~W.~Bennett {\it et al.}  [Muon G-2 Collaboration],
  Phys.\ Rev.\ D {\bf 73}, 072003 (2006)
  [hep-ex/0602035];
  F.~Jegerlehner and A.~Nyffeler,
  Phys.\ Rept.\  {\bf 477}, 1 (2009)
  [arXiv:0902.3360 [hep-ph]].
  M.~Davier, A.~Hoecker, B.~Malaescu and Z.~Zhang,
  Eur.\ Phys.\ J.\ C {\bf 71}, 1515 (2011)
  [Erratum-ibid.\ C {\bf 72}, 1874 (2012)]
  [arXiv:1010.4180 [hep-ph]];
  K.~Hagiwara, R.~Liao, A.~D.~Martin, D.~Nomura and T.~Teubner,
  J.\ Phys.\ G {\bf 38}, 085003 (2011)
  [arXiv:1105.3149 [hep-ph]].

\bibitem{MuonMagMomentNewPhysics}
  H.~Davoudiasl, H.~-S.~Lee and W.~J.~Marciano,
  Phys.\ Rev.\ Lett.\  {\bf 109}, 031802 (2012)
  [arXiv:1205.2709 [hep-ph]];
  S.~Kanemitsu and K.~Tobe,
  Phys.\ Rev.\ D {\bf 86}, 095025 (2012)
  [arXiv:1207.1313 [hep-ph]].

\bibitem{4thGenQuarks}
  M.~M.~H.~Luk,
  arXiv:1110.3246 [hep-ex];
  S.~Chatrchyan {\it et al.}  [CMS Collaboration],
  Phys.\ Lett.\ B {\bf 716}, 103 (2012)
  [arXiv:1203.5410 [hep-ex]];
  G.~Aad {\it et al.}  [ATLAS Collaboration],
  Phys.\ Rev.\ Lett.\  {\bf 109}, 032001 (2012)
  [arXiv:1202.6540 [hep-ex]];
  S.~Chatrchyan {\it et al.}  [CMS Collaboration],
  JHEP {\bf 1205}, 123 (2012)
  [arXiv:1204.1088 [hep-ex]].

\bibitem{3body}
  T.~Toma,
  Phys.\  Rev.\  Lett.\  {\bf 111}, 091301 (2013)
  [arXiv:1307.6181 [hep-ph]];
   F.~Giacchino, L.~Lopez-Honorez and M.~H.~G.~Tytgat,
  arXiv:1307.6480 [hep-ph].

\bibitem{Kumar:2013iva}
  J.~Kumar and D.~Marfatia,
  arXiv:1305.1611 [hep-ph].

\bibitem{Cheung:2009fc}
  K.~Cheung, O.~C.~W.~Kong and J.~S.~Lee,
  JHEP {\bf 0906}, 020 (2009)
  [arXiv:0904.4352 [hep-ph]].

\bibitem{Jung:2008it}
  S.~Jung and J.~D.~Wells,
  Phys.\ Rev.\ D {\bf 80}, 015009 (2009)
  [arXiv:0811.4140 [hep-ph]].

  \bibitem{muon}
  http://pdg.lbl.gov/2011/reviews/rpp2011-rev-g-2-muon-anom-mag-moment.pdf

  \bibitem{mm}
  http://pdg.lbl.gov/2012/reviews/rpp2012-rev-g-2-muon-anom-mag-moment.pdf

  \bibitem{electron}
  http://pdg.lbl.gov/2012/listings/rpp2012-list-electron.pdf

\bibitem{Moyotl:2011yv}
  A.~Moyotl, A.~Rosado and G.~Tavares-Velasco,
  Phys.\ Rev.\ D {\bf 84}, 073010 (2011)
  [arXiv:1109.4890 [hep-ph]].

  \bibitem{pdg}
 C.~Amsler et al. (Particle Data Group),
 Phys.\ Lett.\ B {\bf 667}, 1 (2008).

\bibitem{Carpenter:2011wb}
  L.~M.~Carpenter,
  arXiv:1110.4895 [hep-ph].


\bibitem{lhcslepton}
  G.~Aad {\it et al.}  [ATLAS Collaboration],
  Phys.\ Lett.\ B {\bf 718}, 879 (2013)
  [arXiv:1208.2884 [hep-ex]];
  CMS PAS-SUS-12-002.

\bibitem{lhcsleptonrh}
ATLAS Collaboration,
ATLAS-CONF-2013-049 (2013).


\bibitem{Buckley:2013sca}
  M.~R.~Buckley, D.~Hooper and J.~Kumar,
  arXiv:1307.3561 [hep-ph].


\end{thebibliography}
\end{document}